\PassOptionsToPackage{unicode}{hyperref}
\PassOptionsToPackage{hyphens}{url}
\PassOptionsToPackage{dvipsnames,svgnames,x11names}{xcolor}
\documentclass[
  12pt,
  a4paper,
]{article}
\usepackage{xcolor}
\usepackage[margin=1in]{geometry}
\usepackage{amsmath,amssymb}
\usepackage{iftex}
\ifPDFTeX
  \usepackage[T1]{fontenc}
  \usepackage[utf8]{inputenc}
  \usepackage{textcomp} 
\else 
  \usepackage{unicode-math} 
  \defaultfontfeatures{Scale=MatchLowercase}
  \defaultfontfeatures[\rmfamily]{Ligatures=TeX,Scale=1}
\fi
\usepackage{lmodern}
\ifPDFTeX\else
\fi
\IfFileExists{upquote.sty}{\usepackage{upquote}}{}
\IfFileExists{microtype.sty}{
  \usepackage[]{microtype}
  \UseMicrotypeSet[protrusion]{basicmath} 
}{}
\usepackage{setspace}
\makeatletter
\@ifundefined{KOMAClassName}{
  \IfFileExists{parskip.sty}{%
    \usepackage{parskip}
  }{
    \setlength{\parindent}{0pt}
    \setlength{\parskip}{6pt plus 2pt minus 1pt}}
}{
  \KOMAoptions{parskip=half}}
\makeatother
\makeatletter
\ifx\paragraph\undefined\else
  \let\oldparagraph\paragraph
  \renewcommand{\paragraph}{
    \@ifstar
      \xxxParagraphStar
      \xxxParagraphNoStar
  }
  \newcommand{\xxxParagraphStar}[1]{\oldparagraph*{#1}\mbox{}}
  \newcommand{\xxxParagraphNoStar}[1]{\oldparagraph{#1}\mbox{}}
\fi
\ifx\subparagraph\undefined\else
  \let\oldsubparagraph\subparagraph
  \renewcommand{\subparagraph}{
    \@ifstar
      \xxxSubParagraphStar
      \xxxSubParagraphNoStar
  }
  \newcommand{\xxxSubParagraphStar}[1]{\oldsubparagraph*{#1}\mbox{}}
  \newcommand{\xxxSubParagraphNoStar}[1]{\oldsubparagraph{#1}\mbox{}}
\fi
\makeatother

\usepackage{longtable,booktabs,array}
\usepackage{calc} 
\usepackage{etoolbox}
\makeatletter
\patchcmd\longtable{\par}{\if@noskipsec\mbox{}\fi\par}{}{}
\makeatother
\IfFileExists{footnotehyper.sty}{\usepackage{footnotehyper}}{\usepackage{footnote}}
\makesavenoteenv{longtable}
\usepackage{graphicx}
\makeatletter
\newsavebox\pandoc@box
\newcommand*\pandocbounded[1]{
  \sbox\pandoc@box{#1}%
  \Gscale@div\@tempa{\textheight}{\dimexpr\ht\pandoc@box+\dp\pandoc@box\relax}%
  \Gscale@div\@tempb{\linewidth}{\wd\pandoc@box}%
  \ifdim\@tempb\p@<\@tempa\p@\let\@tempa\@tempb\fi
  \ifdim\@tempa\p@<\p@\scalebox{\@tempa}{\usebox\pandoc@box}%
  \else\usebox{\pandoc@box}%
  \fi%
}
\def\fps@figure{htbp}
\makeatother

\usepackage[style=authoryear,]{biblatex}
\makeatletter
\@ifpackageloaded{caption}{}{\usepackage{caption}}
\AtBeginDocument{%
\ifdefined\contentsname
  \renewcommand*\contentsname{Table of contents}
\else
  \newcommand\contentsname{Table of contents}
\fi
\ifdefined\listfigurename
  \renewcommand*\listfigurename{List of Figures}
\else
  \newcommand\listfigurename{List of Figures}
\fi
\ifdefined\listtablename
  \renewcommand*\listtablename{List of Tables}
\else
  \newcommand\listtablename{List of Tables}
\fi
\ifdefined\figurename
  \renewcommand*\figurename{Figure}
\else
  \newcommand\figurename{Figure}
\fi
\ifdefined\tablename
  \renewcommand*\tablename{Table}
\else
  \newcommand\tablename{Table}
\fi
}
\@ifpackageloaded{float}{}{\usepackage{float}}
\floatstyle{ruled}
\@ifundefined{c@chapter}{\newfloat{codelisting}{h}{lop}}{\newfloat{codelisting}{h}{lop}[chapter]}
\floatname{codelisting}{Listing}

\makeatother
\makeatletter
\@ifpackageloaded{caption}{}{\usepackage{caption}}
\@ifpackageloaded{subcaption}{}{\usepackage{subcaption}}
\makeatother
\usepackage{bookmark}
\IfFileExists{xurl.sty}{\usepackage{xurl}}{} 
\hypersetup{
  colorlinks=true,
  linkcolor={blue},
  filecolor={Maroon},
  citecolor={Blue},
  urlcolor={Blue},
  pdfcreator={LaTeX via pandoc}}

\author{}
\date{}
\begin{document}

\setstretch{1.5}
\begin{center}
{\LARGE \textbf{tsspiral: An R Package to Create Circular Versions of Time Series Plots}}\\
Thiyanga S. Talagala \\
Department of Statistics, Faculty of Applied Sciences, \\
University of Sri Jayewardenepura, Sri Lanka \\
ttalagala@sjp.ac.lk
\end{center}

\section*{Abstract}\label{abstract}
\addcontentsline{toc}{section}{Abstract}

\texttt{tsspiral} is an open-source R package that introduces a circular
approach to time series visualisation using polar coordinates. We refer
to this circular representation of time series data as a spiral.
\texttt{tsspiral} provides a collection of \texttt{ggplot2} geometries
for creating a spiral representation of time series data. In our
approach, we arrange observations around a circle. Each year is
represented by a circle, with the angular position corresponding to the
day of the year, and successive years appear as concentric rings. This
arrangement facilitates the exploration of periodicity, seasonality, and
recurring temporal patterns, particularly in long time series. The
package also supports temporal hierarchical visualisations to compare
patterns across time scales. By bringing polar-coordinate-based
visualisation into routine time series analysis, \texttt{tsspiral}
offers a practical tool for exploring temporal patterns and
communicating complex time-dependent structures. Rather than treating
spiral plots as standalone graphics, our approach brings them into the
grammar of graphics framework, enabling users to explore temporal
patterns through a broader range of visual encodings.

\textbf{keywords:} missing values, forecasting, daily series,
visulisation, seasonal patterns, polar coordinates

\section{Statement of need}\label{statement-of-need}

Time series visualization is an essential step in time series analysis,
forecasting, imputation, anomaly detection etc. Time series data are
commonly visualized using Cartesian coordinate plane, where time is
represented along the horizontal axis (\(x\)) and observed values along
the vertical axis (\(y\)). Although conventional time series plots are
effective for identifying important temporal features, such as trends,
seasonal patterns, sudden shifts, and irregular fluctuations,
identifying these features can become challenging with long time series,
particularly daily observations spanning several years or decades. As
the length of the series increases, the visual representation becomes
increasingly dense, making it difficult to recognize recurring patterns,
compare corresponding periods across years and identify changes in
seasonal behavior.

Circular representations of time series provide an alternative
visualisation in which time is mapped to angular position and successive
years are arranged as concentric rings. This representation facilitates
visual comparison of corresponding time periods across years, making
recurrent patterns, unusual events, temporal changes, and missing data
gaps easier to explore. Furthermore, conventional time series plots make
it difficult to visualise nested temporal structures and compare
seasonal patterns across different levels of the temporal hierarchy. In
addition, traditional time series plots can make categorical time series
data hard to visualise. Spiral time series plots help address these
limitations.

Spiral plots support visualising long time series and better identifying
periodic structures in the data
\autocite{weber2001visualizing,rakuschek2026guided}. John W. Tukey
emphasized, ``Exploratory data analysis is an attitude, a state of
flexibility, a willingness to look for those things that we believe are
not there, as well as those we believe to be
there''\autocite{tukey1977exploratory}. Guided by this philosophy,
exploring alternative visualizations of time series data encourages us
to move beyond familiar representations and discover patterns that might
otherwise remain hidden. Spiral layout has two major advantages for
visualizing time series data: i) able to visualize data with long axes,
which greatly improves the resolution of visualization and space; ii)
efficient for time series data to reveal periodic patterns
\autocite{spiralize}.

Since most statistical software packages readily support line plots,
their convenience and widespread availability have made line plots the
conventional choice for visualizing time series. As a result, other
forms of visualization have received comparatively less attention,
despite their potential to reveal recurring and circular patterns in
time series data. Furthermore, creating spiral time series plots often
requires custom code and careful handling of time series data.
\texttt{tsspiral} fills these gaps.

\section{State of the field}\label{state-of-the-field}

\textcite{carlis1998interactive} is one of the early contributions that
introduced arranging data in a spiral. The authors mentioned that
spirals provide the user with easy visual cues to both serial and
periodic aspects of the data. \textcite{weber2001visualizing} proposes
spiral graphs for visualizing long time-series data sets and identifying
periodic structures. It also discusses visualizing nominal and
quantitative time series data. The R package
\texttt{spiralize} \autocite{spiralize} provides tools for visualizing
time series data using spiral layouts. Our package introduces a new set
of geoms that integrate with \texttt{ggplot2} \autocite{ggplot2},
extending the grammar of graphics to spiral-based visualization. This
integration lets users build spiral plots with a familiar, flexible
syntax while retaining the layered approach to data visualisation that
makes \texttt{ggplot2} powerful.

\section{Software design}\label{software-design}

The central idea of \texttt{tsspiral} is to transform a linear time
series into a sequence of annual circular representations. Each year
occupies a separate ring. Each year has 365 segments, while a leap year
has 366. Concentric rings represent successive years. The angular
coordinate represents an observation's position within its corresponding
calendar year.

Let $t$ denote a date and \(y_t\) its associated observation. The number
of days in year $X$ is defined as

\[
D_X =
\begin{cases}
366, & \text{if }X\text{ is a leap year},\\
365, & \text{otherwise}.
\end{cases}
\]

Each annual ring is divided into $D_X$ equal angular segments, where
$D_X$ represents the number of days in calendar year $X$. Each segment
corresponds to one calendar day. We places January 1 at the top of each
annual ring, with subsequent days progressing clockwise. The angular
width of each segment is therefore

\[
\Delta\theta_X = \frac{2\pi}{D_X}.
\]

Consequently, each annual ring completes exactly one full revolution.
The radial coordinate identifies the year, with each successive year
assigned to a new concentric ring. It can be expressed as \[
r_t = r_0 + k\Delta r,
\]

where $r_0$ is the radius of the innermost ring, $\Delta r$ is the
radial distance between consecutive annual rings, and \(k\) is the
zero-based index of the year relative to the starting year.

For a date \(t\) belonging to calendar year \(X\), let \(d_t\) denote
its day-of-year index, where January 1 corresponds to \(d_{t}=0\). The
angular coordinate is then defined as \[
\theta_t = \frac{2\pi d_{t}}{D_X}.
\]

Thus, the polar coordinates of an observation \(y_t\) are

\[
(r_t,\theta_t)=
\left(
r_0+k\Delta r,\,
\frac{2\pi d_t}{D_X}
\right).
\] Figure 1 shows the graphical representation of the construction of
spiral plots using the \texttt{tsspiral} package.

\begin{figure}

\centering{

\includegraphics[width=1\linewidth,height=\textheight,keepaspectratio]{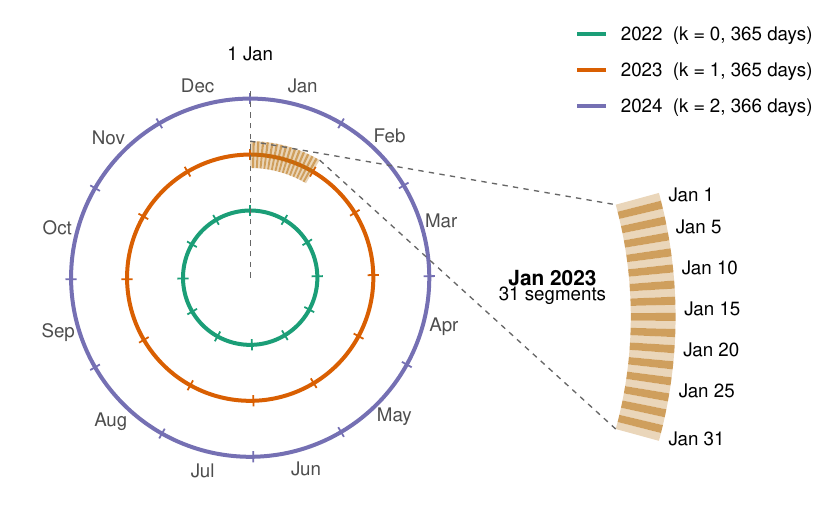}

}

\caption{\label{fig-f1}Graphical representation of the construction of
spiral plots using the tsspiral package. Each year is represented by a
circular ring, with the angular position corresponding to the day of the
year. Successive years are arranged as concentric rings, allowing
examination of temporal patterns across years.}

\end{figure}%

The package provides three main geometries. They are listed as follows:

\begin{enumerate}
\def\labelenumi{\roman{enumi}.}
\item
  Spiral time series geometry (\texttt{geom\_tsspiral()}): The primary
  geometry transforms time series observations into a spiral layout.
  Each year is represented by a separate ring, and observations are
  positioned according to their temporal locations. The geometry
  supports visualizing daily, weekly, monthly, quarterly and annual
  observations. For lower-frequency observations, the corresponding
  temporal intervals are expanded to daily positions, allowing
  observations to occupy the appropriate calendar regions. For example,
  for monthly data, the January value is represented by colouring the
  segments corresponding to January 1 to January 31 using the January
  value. The \texttt{ring\_spacing} argument controls the separation
  between annual rings. Calendar labels are displayed around the
  outermost ring.
\item
  Spiral bar geometry (\texttt{geom\_tsspiralbar()}): This geometry
  represents time series observations using bars instead of coloring the
  segment according to its value. The angular position identifies the
  temporal location, while the radial extent of each bar represents the
  magnitude of the corresponding observation. This representation is
  useful when comparing the magnitude and seasonal distribution of
  observations across multiple years.
\item
  Missing value representation geometry (\texttt{geom\_tsspiralmiss()}):
  The missing segments of the time series are represented by coloring
  the missing segment in black. It enables users to identify periods of
  incomplete data collection and examine whether missingness is
  concentrated within particular calendar periods or years.
\end{enumerate}

In addition to these geometries, \texttt{tsspiral} provides two further
functions:

\begin{itemize}
\item
  \texttt{temporal\_hierarchical()}: represents daily, weekly, monthly,
  quarterly and annual summaries using aligned concentric rings. This
  enables the examination of temporal hierarchical patterns at multiple
  resolutions within a common coordinate system.
\item
  \texttt{calendar\_hourly\_polar()}: creates a monthly calendar in
  which each day contains a 24-hour polar bar chart, allowing intra-day
  patterns to be examined while retaining calendar context.
\end{itemize}

Together, these functionalities provide a collection of complementary
approaches for exploring temporal structures at different resolutions.

\section{\texorpdfstring{Exploring time series patterns with
\texttt{tsspiral}}{Exploring time series patterns with tsspiral}}\label{exploring-time-series-patterns-with-tsspiral}

Figure 2 shows an example illustration. Figure 2 (a) shows the time
series plot of daily tourist arrivals to Sri Lanka from 2023 January 1
to 2025 December 31. The corresponding data set is available in the
\texttt{tsspiral} R package. This time series plot helps us to
understand the overall behavior of the data across the full time span.
We can see a clear upward trend and a seasonal behavior in the data. But
through the line time series plot, it is difficult to answer the
questions: i) 1. When is Sri Lanka busiest? ii). When is it quiet? iii).
Are there predictable peaks and troughs throughout the year? iv). Which
months or seasons attract more visitors? v). Do certain days or periods
consistently rank among the busiest? vi). Are there recurring days with
lower arrivals that could be used for maintenance, conservation, or
quiet operations? vii) . How do recurring seasonal patterns differ
across years---for example, between January 2023, January 2024, and
January 2025?. The same time series plotted using the
\texttt{geom\_tsspiral} are shown in Figure 2-b and Figure 2-c.~Figure
2-b shows the numerical values, and Figure 2-c shows the categorised
values. The spiral version makes it easy to answer our question. For
example, we observe peaks every year from July to August and from late
November to January of the next year. Arrivals are low in May every
year, which we can use for maintenance, conservation, or quiet
operations.

\begin{figure}[H]

{\centering \includegraphics[width=1\linewidth,height=0.5\textheight]{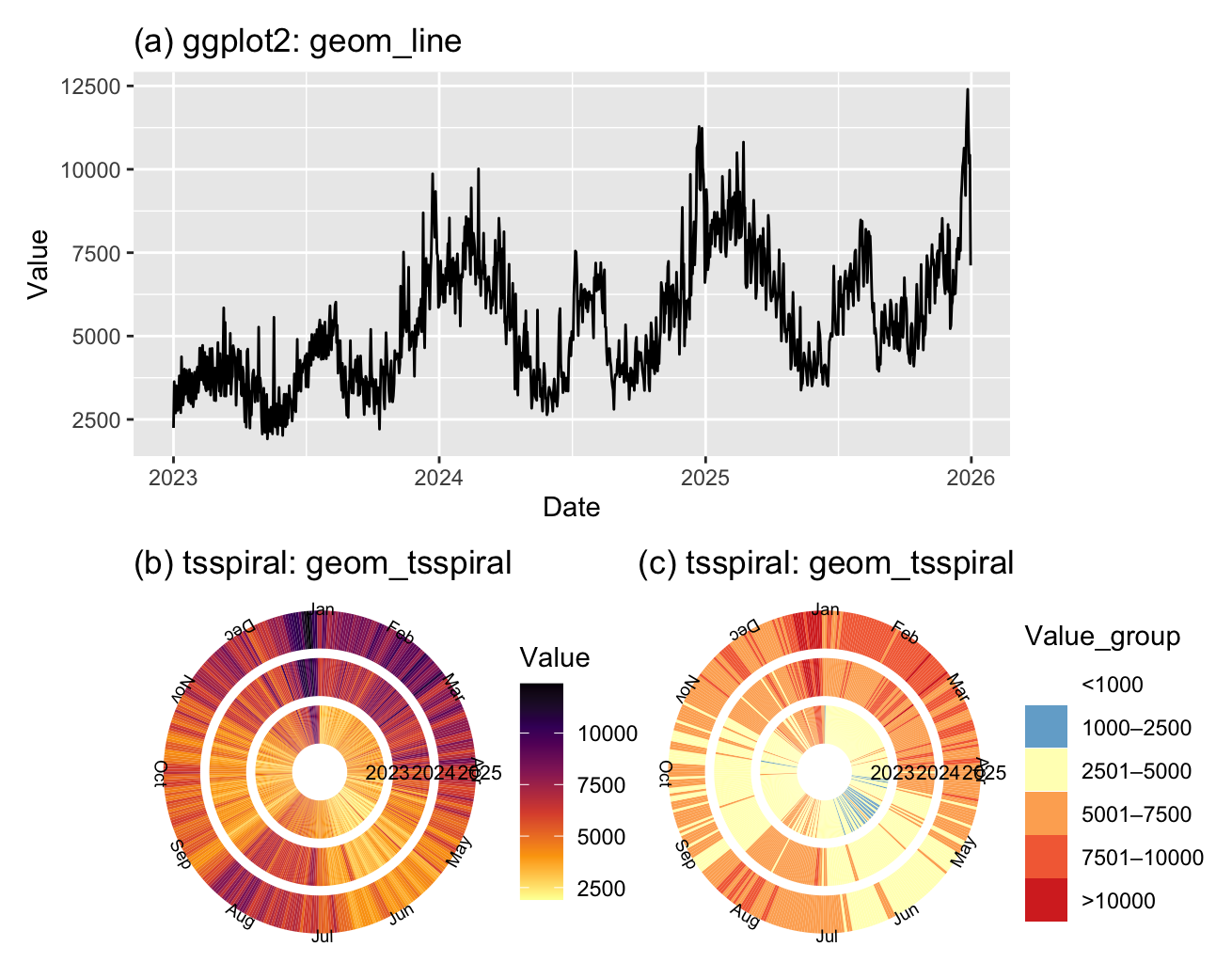}

}

\caption{Illustration example using Sri Lanka daily tourist arrival
data. (a) time series plot using geom~point, (b) spiral version of plot
representing numeric values, (c) spiral version of the plot representing
categorical values. Every year tourist mostly arrive in July- August and
from December to January of the next year.}

\end{figure}%

\begin{figure}[H]

{\centering \includegraphics[width=0.9\linewidth,height=0.5\textheight]{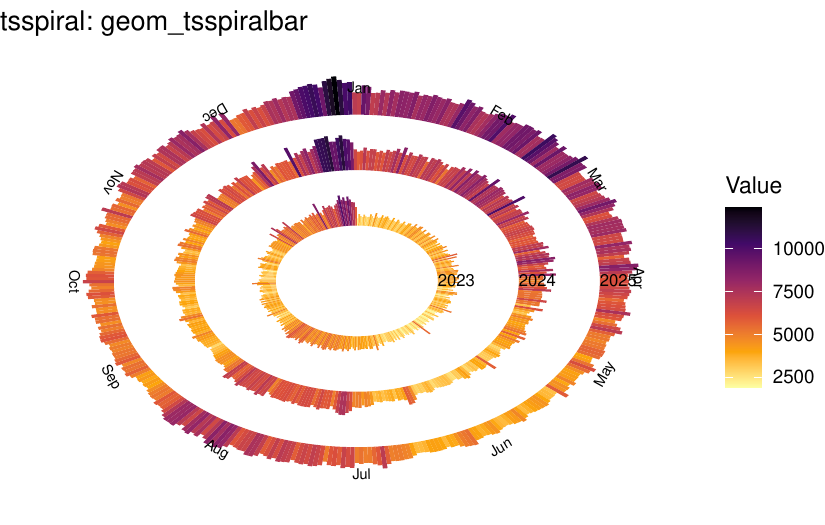}

}

\caption{Illustration example representing Sri Lanka daily tourist
arrival data using geom~tsspiralbar. The outer circle bar height being
greater than the inner circle bars indicates the upward trend in the
series. Furthermore, we can clearly see that December each year records
the highest number of arrivals compared to other months. There is a
significant upward trend in 2024 compared to 2023.}

\end{figure}%

Figure 3 shows the daily tourist arrivals to Sri Lanka using the
\texttt{geom\_tsspiralbar}. The bar height and colour are coded
according to the observed values. The outer circle bar height being
greater than the inner circle bars indicates the upward trend in the
series. Furthermore, we can clearly see that December each year records
the highest number of arrivals compared to other months. There is a
significant upward trend in 2024 compared to 2023.

The temporal\_hierarchical() function can be used to visualise temporal
patterns at different levels of temporal aggregation. Users can specify
an aggregation function through the fun argument, providing flexibility
in how observations are summarised. For example, max can be used to
display the maximum value, mean to represent the average value, and sum
to show the total value within each temporal aggregation period. This
flexibility allows users to explore different aspects of the data and
identify temporal patterns according to their analytical objectives.
Figure 4 represents temporal hierarchical visualisation of the 2023
daily tourist arrivals to Sri Lanka using the maximum value as the
aggregation function. The maximum occurs in the fourth quarter, in
December, and corresponds to the last weeks of the year.

\begin{figure}[H]

{\centering \includegraphics[width=1.3\linewidth,height=0.5\textheight]{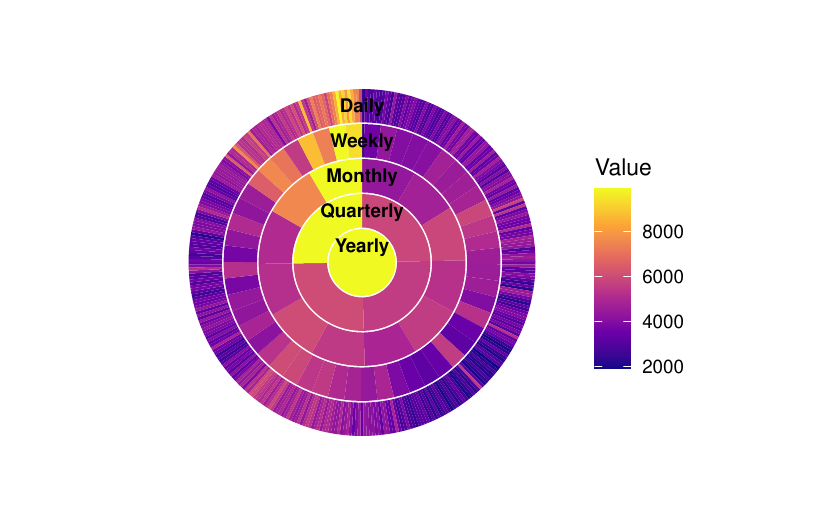}

}

\caption{Temporal hierarchical visualisation of 2023 daily tourist
arrivals to Sri Lanka, using the maximum value as the aggregation
function. The maximum occurs in the fourth quarter, in December, and
corresponds to the last weeks of the year.}

\end{figure}%

Figure 5 illustrates the \texttt{geom\_tsspiralmiss} function in the
package. To demonstrate its functionality, we deliberately remove
selected rows from the original tourist arrivals to Sri Lanka dataset, creating gaps in the time
series. The resulting visualisation highlights missing time points and
demonstrates how \texttt{geom\_tsspiralmiss} represents gaps in temporal
coverage.

\begin{figure}[H]

{\centering \includegraphics[width=0.9\linewidth,height=0.6\textheight]{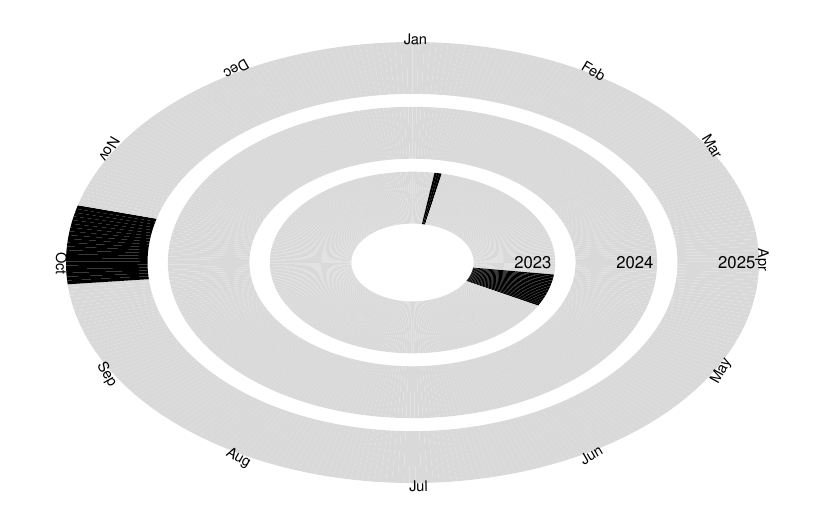}

}

\caption{An illustrative example \texttt{geom\_tsspiralmiss} function in
the package. The missing data sections are denoted in black.}

\end{figure}%

Inspired by the work of \textcite{wang2020calendar}, we use
calendar-based plots to represent hourly time series. Figure 6
illustrates the approach using simulated hourly data. Each panel
represents a day, with the spiral plot within each panel showing the
distribution of hourly observations throughout that day.

\begin{figure}[H]

{\centering \includegraphics[width=1\linewidth,height=0.6\textheight]{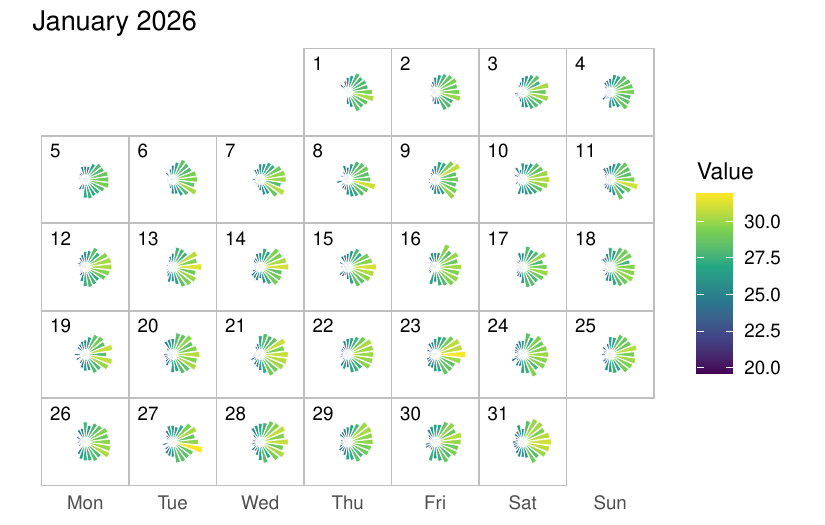}

}

\caption{An illustrative example \texttt{calendar\_hourly\_polar}
function in the package.}

\end{figure}%

\section{Research impact statement}\label{research-impact-statement}

\texttt{tsspiral} expands the ways researchers can explore time series
through a circular shape. By representing time series in polar
coordinates, the package provides an alternative way to explore temporal
patterns, recurring cycles and changes over time. Furthermore, this
allows visualisation of temporal-hierarchical time series. By
aggregating time series under different temporal scales from years to
quarters, months, weeks, and days, spiral visualizations can reveal
recurring seasonal patterns and relationships between different temporal
scales. This allows researchers to explore patterns in the time series
within and across years, quarters, months, or other time intervals. The
package \texttt{tsspiral} provides one stop shop for visualizing time
series series data in circular arrangement.
\textcite{rakuschek2026guided} pointed out that circular time-series
plots have both advantages and disadvantages. Hence,
\textcite{rakuschek2026guided} suggested using both line and circular
versions of the plots to better understand the data, rather than using
circular plots as a replacement for line plots.

\subsubsection*{AI usage disclosure}\label{ai-usage-disclosure}
\addcontentsline{toc}{subsubsection}{AI usage disclosure}

The software project presented in this paper accepts contributions that
have been created with the help of AI tools. ChatGPT was used to assist
with code efficiency improvement and documentation roxygen files
writing. Contributors are responsible for reviewing all AI-generated
content and ensuring its correctness and maintainability. AI tools were
used to improve the readability of this paper and to check spelling and
grammar.

\subsubsection*{Reproducibility}\label{reproducibility}
\addcontentsline{toc}{subsubsection}{Reproducibility}

The R package required to reproduce the results presented in this paper
is available from CRAN \autocite{tsspiral} and GitHub at
\url{https://github.com/thiyangt/tsspiral}. The complete R scripts and
supporting code used to generate the results, figures, and tables are
available at \url{https://github.com/thiyangt/tsspiral-paper}. These
resources enable readers to reproduce the analyses and visualizations
presented in the paper.

\printbibliography[title=References]

\end{document}